# First Steps in Relational Lattice

MARSHALL SPIGHT
Marshall.Spight@gmai1.com
VADIM TROPASHKO
Vadim.Tropashko@orcl.com

___________________________________________________________________

Relational lattice reduces the set of six classic relational algebra operators to two binary lattice operations: natural join and inner union. We give an introduction to this theory with emphasis on formal algebraic laws. New results include Spight distributivity criteria and its applications to query transformations.

___________________________________________________________________

## 1. INTRODUCTION

There are many reasons why classic relational algebra is unsatisfactory from mathematical perspective. First, the set of six basic operations is quite large. With so many operators it is challenging to establish relational algebra axiomatization. Without such axiomatization the database implementation field is a collection of ad-hock methods. Witness the eight query rewrite rules listed in section 4.4 of the Alice book ([1]). How can we be sure that no rule is missing?

Second, the operations are syntactically unattractive. The elements of the algebra are relations, and yet some operations like projection and selection take an additional parameter, which is outside of the realm of the relation objects. Some operations like union can't be applied to any pair of relations. The explicit renaming operation is like nothing else in mathematics, where renaming variables has never been a big deal.

Relational lattice [2] reduces the number of basic operations to just two. The first operation is the familiar *natural join*

$$A(x,y) \bowtie B(y,z) \stackrel{def}{=} \{(x,y,z) \mid (x,y) \in A \land (y,z) \in B \}$$

The second operation is the *inner union*

$$A(x,y) \unlhd B(y,z) \stackrel{def}{=} \{(y) \mid \exists x\ (x,y) \in A \lor \exists z\ (y,z) \in B \}$$

Natural join and inner union honor the *lattice axioms*

**Idempotent laws:**

$A \bowtie A = A \qquad\qquad\qquad A \unlhd A = A$

**Commutative laws:**

$A \bowtie B = B \bowtie A$  $\qquad A \veebar B = B \veebar A$

**Associative laws:**

$A \bowtie (B \bowtie C) = (A \bowtie B) \bowtie C$  $\qquad A \veebar (B \veebar C) = (A \veebar B) \veebar C$

**Absorption laws:**

$A \veebar (A \bowtie B) = A$  $\qquad A \bowtie (A \veebar B) = A$

hence the name – *Relational Lattice*.

Extending the classic relational algebra union operation to be applicable to any pair of relations is not particularly original. Date and Darwen define the OR operation in [3] as

$$A(x,y) \text{ OR } B(y,z) = \{(x,y,z) \mid (x,y) \in A \land \exists z \in Z\} \cup \{(x,y,z) \mid \exists x \in X \land (y,z) \in B\}$$

where the X and Z are the domains for x and z, correspondingly. Apparently, variables ranging over the whole domain raise the *domain safety* issues.

Next, the OR operation is idempotent, commutative, and associative. Formally, the OR algebra is an *upper semilattice*. Likewise, the AND operation[1] defines the *lower semilattice*. Any semilattice defines a partial order relation. In Date & Darwen algebra we have the two partial orders:

$$A \preccurlyeq B \quad \text{iff} \quad A \text{ OR } B = A$$

$$A \precsim B \quad \text{iff} \quad A \text{ AND } B = B$$

It is the absorption law that ensures that both partial orders coincide. Without the absorption law the upper semilattice is *incompatible* with the lower semilattice. This algebra is studied in more depth in [4].

Finally, Date & Darwen algebra lacks the reduction capabilities that we study in the next section.

2. REDUCTION

Relational model draws a demarcation line between predicates and relations. The former are *extensional*, the while latter are *intensional*. We blur the distinction between the two, which implies that we have to consider infinite relations as first class citizens. Arguably, the most prominent example of such a relation is the equality predicate

$$E(x,y) = \{(x,y) \mid x = y\}$$

Let's introduce a couple of convenient abbreviations. A predicate, for example $\{(x,y) \mid x \leq y\}$, will be written as $\lceil x \leq y \rceil$. An empty relation $\{(x,y) \mid x = 0 \land x \neq 0 \land y = 0 \land y \neq 0\}$, where 0 is understood to be any element in the appropriate domain, will be written as ⌈xy⌉. Now we have syntax to express the classic relational algebra operations in terms of natural join and inner union.

---
[1] The AND operation in Date & Darwen algebra is the natural join as well.

**Selection:** $\sigma_{x>y} A(x,y,z) = A(x,y,z) \bowtie \lceil x>y \rceil$.

Reducing selection to join is intriguing from query optimization perspective. Classic optimization method represents query execution as a tree whose leaves are database relations and branch nodes are operations. Relational algebra rewriting rules, such as join commutativity and associativity, allow transforming query tree to equivalent forms. Then, each execution tree is evaluated in terms of cost, and the one with the lowest cost is selected. In our case we have the two alternative join order permutations[2]:

1. Consider A as a relation on the outer side of the join. Fetch first tuple. Find all the matching tuples from the ⌈x>y⌉ relation. Fetch the next tuple from A…
2. Consider the ⌈x>y⌉ as the leading relation in the join order. Find all the matching tuples from A. Fetch the next tuple from ⌈x>y⌉ …

Clearly, with the infinite domain the relation ⌈x>y⌉ is infinite; therefore, the cost of the second query evaluation strategy is infinite. Yet both methods are legitimate. As soon as we have a finite predicate, say ⌈x=1 ∧ y=0⌉ the second method might become viable. The decision is cost based!

**Projection**: $\pi_y A(x,y) = A(x,y) ⨉ \lceil y \rceil$

**Renaming**: $\rho_{y \to z} A(x,y) = \lceil xz \rceil ⨉ ( A(x,y) \bowtie \lceil y=z \rceil )$

The *set difference* is the only operation that escapes direct representation in terms of natural join and inner union. This setback is not critical for further development, however. One can approach the issue with equational definition in mind, mimicking the way the minus operation is introduced in arithmetics. Given the relations A(z) and B(z), the set difference A \ B is a relation X that satisfies the following system of equations:

$$X \bowtie B = \lceil z \rceil$$
$$X ⨉ B = A$$

This construction has to be equipped with an existence and uniqueness condition, of course.

Alternatively, the set difference A \ B can be expressed via relational division S / R in two steps

$$C = \rho_{z \to y} ( A \bowtie B )$$
$$A \setminus B = ( \sigma_{z \ne y} A \bowtie C ) / C$$

Relational division is a prototypical example of a *set join*. It will be further explored in the quantifiers section.

---

[2] For simplicity let's assume that the *nested loops join* is the only physical join execution method available

## 3. PARTIAL ORDER AND SPECIAL ELEMENTS

Any lattice is a partially ordered set with the order $\leq$ defined as

$$A \leq B \quad \text{iff} \quad B = A \bowtie B$$

or, symmetrically,

$$A \leq B \quad \text{iff} \quad A = A \talloblong B$$

*Hasse diagram* of quite small relational lattice is shown on figure 1.

With partial order we can go on defining the lattice greatest element 10, and the least element 01. Informally, 10 is an empty relation with the header combining all the possible attribute names. Symmetrically, 01 is a relation with no attributes and some nonempty content. All the tuples that have no attributes are equivalent. Therefore, a nonempty relation with no attributes has to have one tuple only.

There are two more special elements. The least element in the sublattice of all the empty relations, which is is the empty relation with no attributes, hence denoted 00. The least element in the sublattice of all nonempty relations with the full set of attributes denoted 11. This is Cartesian product of all the domain relations, or in other words, the celebrated *universal relation*.

| 00 | no attributes | no tuples | the least element among all the header relations |
|----|---------------|-----------|---------------------------------------------------|
| 01 | no attributes | all tuples | the least element in the lattice |
| 10 | all attributes | no tuples | the greatest element in the lattice |
| 11 | all attributes | all tuples | the universal relation |

With new symbols we can use a less sloppy language. Instead of saying "relation A is empty" we write $A \geq 00$. Likewise, instead of "relation A has no attributes" we write $A \leq 00$.

Natural join and inner union of the 00 and 11 elements with an arbitrary relation A produces the following results[3]

- $A \bowtie 00$ is the empty relation with the same header as A. Informally, $A \bowtie 00$ is the set of all the attributes of A. Formally, joining by 00 is lattice homomorphism into Boolean algebra of relation headers.

- $A \talloblong 00 =$ if $A \geq 00$ then 00 else 01. Formally, unioning by 00 is lattice homomorphism into Boolean algebra {00, 01}.

---

[3] Natural join and inner union with 01 and 10 is not interesting (why?)

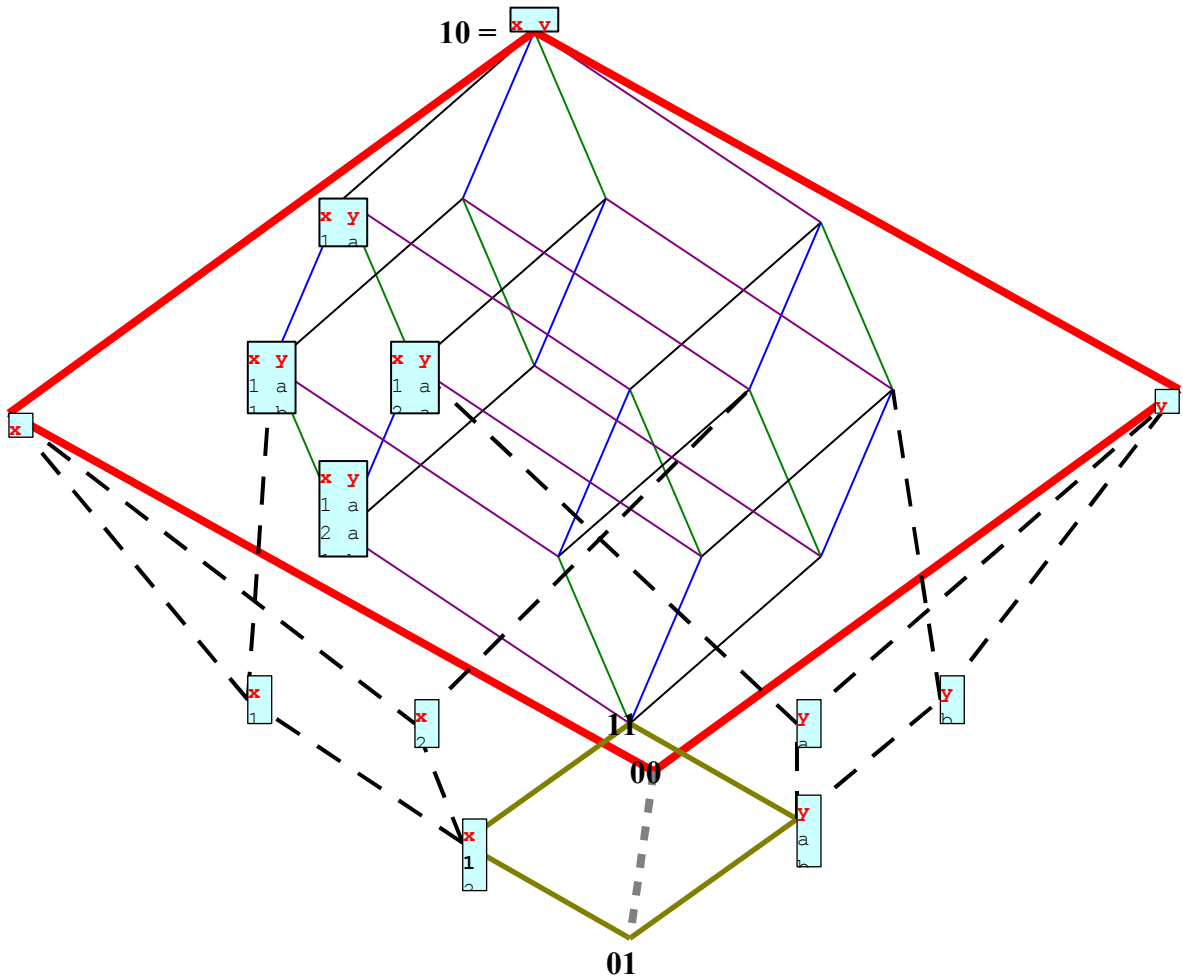

*Figure 1*: A lattice of relations generated out of finite domains x∈{1,2} and y∈{a,b}. For simplicity 10 is identified with ⌈xy⌉, and 11 is identified with {(x,y) | x∈{1,2}, y∈{a,b}}. Six important sublattices of this lattice are highlighted on figures 2 and 3.

- A ⋈ 11 is the A joined with all possible domain relations. Informally, A ⋈ 11 is the content of A stripped off its header. Formally, joining by 11 is lattice homomorphism into Boolean algebra of all subsets of the universal relation.

- A ⧖ 11 is Cartesian product of the domains corresponding to the attributes of A. Formally, unioning by 11 is lattice homomorphism into Boolean algebra of all domains.

Here is couple more translations into the lattice language. "Relation A has a subset of attributes of B" ⇒ A ⋈ 00 ≤ B ⋈ 00. "Relations A and B have the same headers and set of tuples of A is a subset of that of B" ⇒ A ⋈ 00 = B ⋈ 00 and A ≥ B.

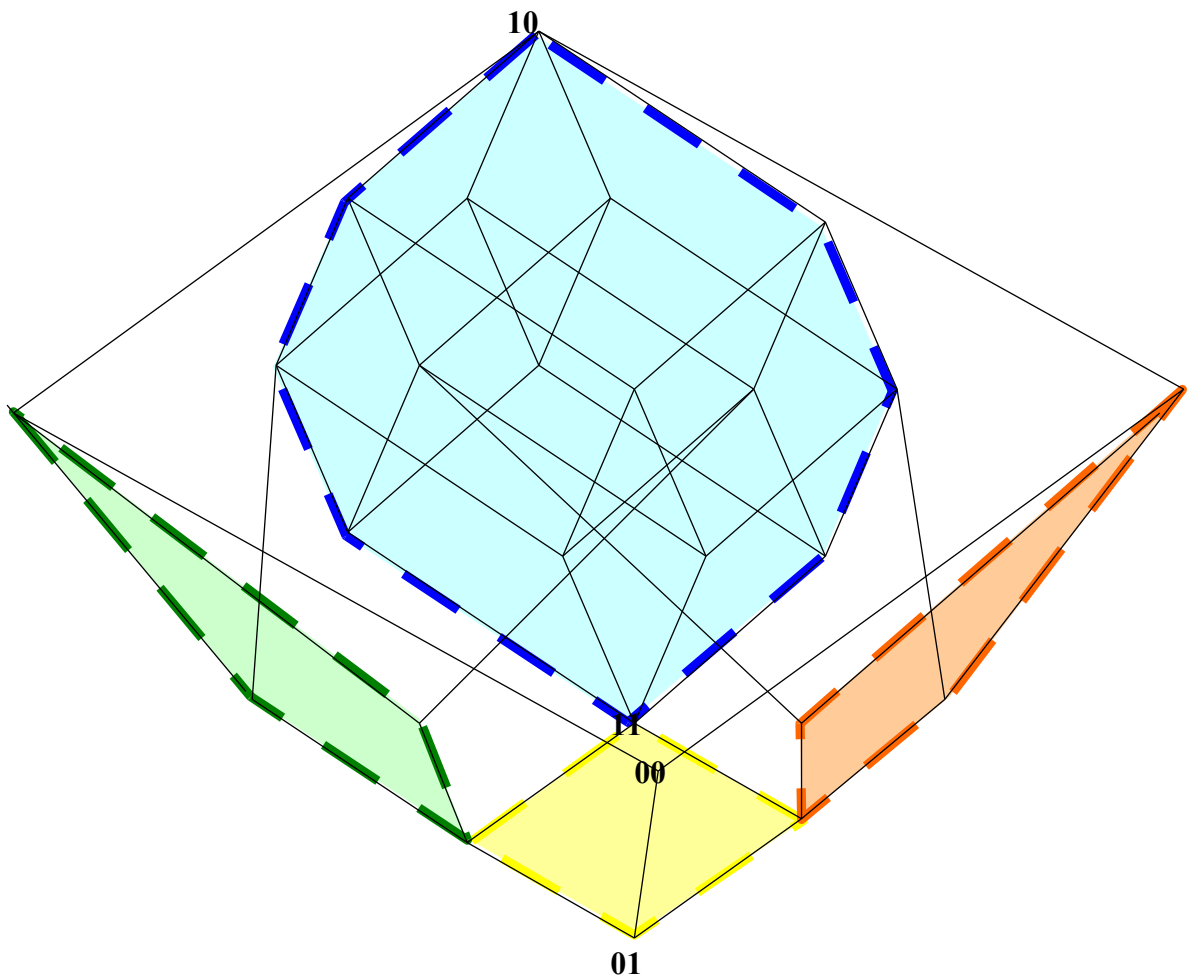

*Figure 2*: Four Boolean algebras as sublattices of the relational lattice. 1. The algebra of all the relations with attributes x and y (blue). 2. The algebra of all the relations with attribute x (green). 3. The algebra of all the relations with attribute y (orange). 4. The algebra with the domains x∈{1,2} and y∈{a,b} atomic elements (yellow).

**Proposition 1.** $\quad A = (A \bowtie 00) \mathbin{\veebar} (A \bowtie 11)$

This fundamental decomposition identity informally asserts that any relation is an inner union of relation's content and header.

## 4. DISTRIBUTIVITY

Algebra of two operations is promising as a basis for practical query transformation engines. Consider "folklore" relational algebra rewrite rules on p.56 of the Alice Book [1]. Most of them easily follow from lattice axioms, for example *push-cross-through-select* can be proved formally in a series of small steps:

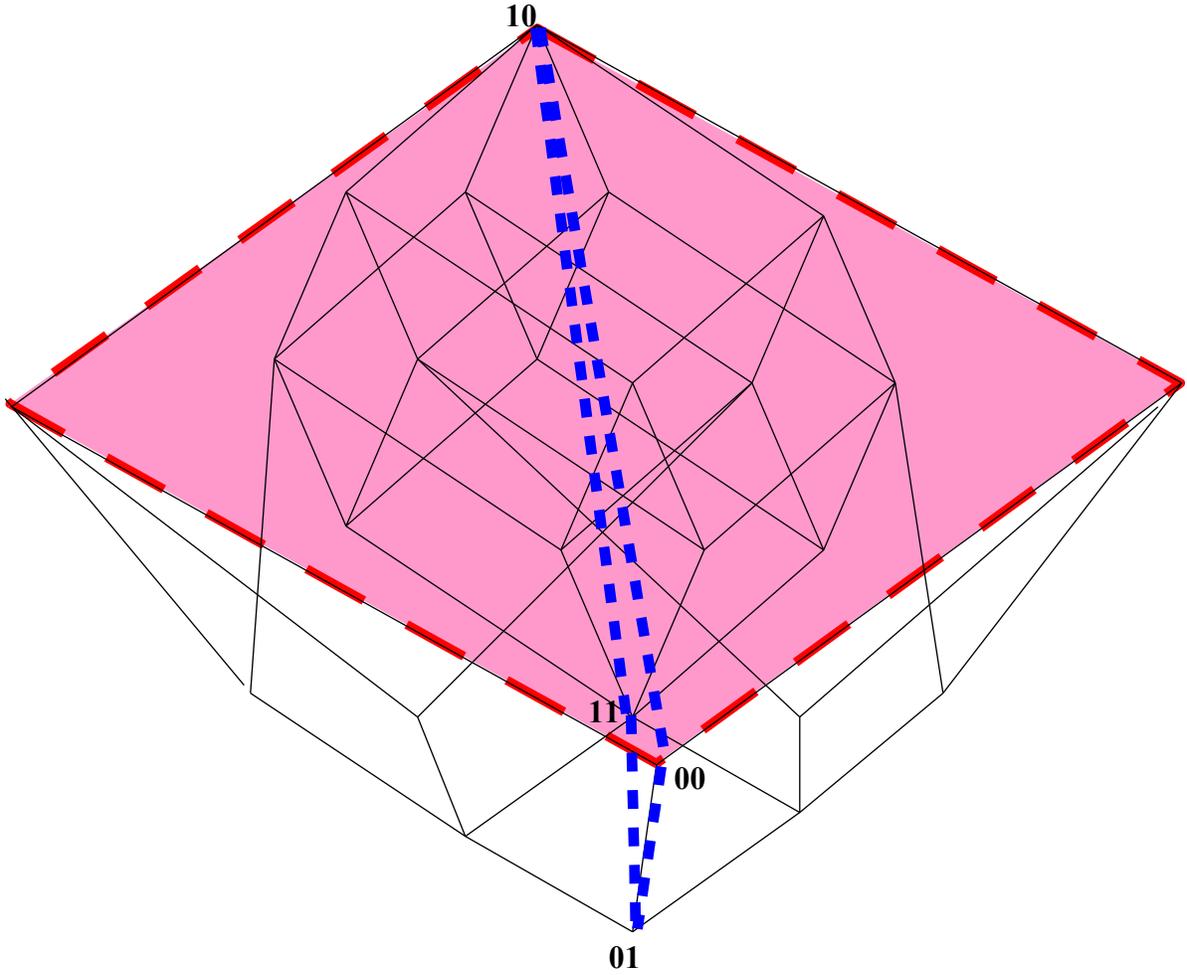

*Figure 3*: Two more sublattices. 1. The algebra of all the empty relations (red). 2. Four element Boolean algebra {10,11,00, 01} (blue).

$$(\sigma_{x=1} A(x,y)) \times B(z) \ =^4 \ (\lceil x=1 \rceil \bowtie A(x,y)) \bowtie B(z) \ =^5$$

$$= \ \lceil x=1 \rceil \bowtie (A(x,y) \bowtie B(z)) \ =^6 \ \sigma_{x=1}(A(x,y) \times B(z))$$

On the other hand, p*ush-select-through-project* can not be worked out without distributivity:

$$\sigma_{x=1}(\pi_{xy} A(x,y,z)) \ =^7 \ \lceil x=1 \rceil \bowtie (\lceil xy \rceil \mathbin{\vcenter{\hbox{$\bowtie$}}} A(x,y)) \ =^8$$

---

[4] rewrite RA expression in lattice terms
[5] apply join associativity
[6] rewrite back to standard RA
[7] rewrite RA expression in lattice terms
[8] distributivity (proved later)

$$= (\lceil x=1 \rceil \bowtie \lceil xy \rceil) \diagX (\lceil x=1 \rceil \bowtie A(x,y,z)) =^9$$

$$= \lceil xy \rceil \diagX (\lceil x=1 \rceil \bowtie A(x,y,z)) =^{10} \pi_{xy} (\sigma_{x=1} A(x,y,z))$$

Likewise, *push-cross-through-project* requires distributivity as well:

$$(\pi_x A(x,y)) \times B(z) =^{11} (\lceil x \rceil \diagX A(x,y)) \bowtie B(z) =^{12}$$

$$= (\lceil x \rceil \bowtie B(z)) \diagX (A(x,y) \bowtie B(z)) =^{13}$$

$$= \lceil xz \rceil \diagX (A(x,y) \bowtie B(z)) =^{14} \pi_{xz} (A(x,y)) \times B(z)$$

The effect of the extra projection column z appearing as a result of the *push-cross-through-project* rewrite has a formal explanation!

Unfortunately, relational lattice is not distributive. Otherwise it would be isomorphic to algebra of sets. Yet, it is easy to see that distributivity holds in at least some circumstances. Figures 2 and 3 highlight a number of sublattices, which are *Boolean algebras*. Evidently, distributivity holds when there is something special about relation headers (e.g all relations have the same header), or there is something special about relation content (e.g. all relations are empty). We'll focus on distributivity criteria that involve relation headers, because a method checking if relation header meets a certain criteria is immediately applicable to the above query transormation applications.

Consider the relations A(x,u,w,t), B(y,u,v,t), and C(z,w,v,t). The attribute x is unique to A, y is unique to B, and z is unique to C. Then, the attribute u is in A and B but not in C, v is in B and C but not in A, and w is in A and C but not in B. Finally, the attribute t is in all the three relations. Adding more attributes to the above relations would change nothing from a distributivity criteria perspective[15].

**Proposition 2**.

1. $A(x,t) \bowtie (B(y,v,t) \diagX C(z,v,t)) = (A(x,t) \bowtie B(y,v,t)) \diagX (A(x,t) \bowtie C(z,v,t))$

In other words, if there is no attributes which are in in A and B but not in C, and also there is no attributes which are in A and C but not in B, then join with A distributes over union of B and C.

---

[9] evaluate join of two constant relations
[10] rewrite back to standard RA
[11] rewrite RA expression in lattice terms
[12] distributivity (established later)
[13] join of empty relation with some relation evaluates to another empty realtion
[14] rewrite back to standard RA
[15] A rigorous (but more awkward) treatment would insist on x, y, z, u, v, w, t being partitions of the set of all the attributes of the relations A, B, and C

2. $A(t) \, \unicode{x2A02} \, (B(y,t) \bowtie C(z,t)) = (A(t) \, \unicode{x2A02} \, B(y,t)) \bowtie (A(t) \, \unicode{x2A02} \, C(z,t))$

This condition is even more restrictive. In addtition to what were required for distributivity of join over union, we demand that there is no attributes which are in B and C but not in A, and also there is no attribute which is unique to A. Then, union with A distributes over join of B and C.

*Proof.* Consider the relations with unrestricted headers. Expanding the left hand side expression

$$A(x,u,w,t) \bowtie (B(y,u,v,t) \, \unicode{x2A02} \, C(z,w,v,t))$$

with set definitions for join and union operations we have

$$\{ (x,u,w,v,t) \mid (x,u,w,t) \in A \wedge (\exists y \exists u \, (y,u,v,t) \in B \vee \exists z \exists w \, (z,w,v,t) \in C ) \}$$

or, equivalently

$$\{ (x,u,w,v,t) \mid ((x,u,w,t) \in A \wedge \exists y \exists u \, (y,u,v,t) \in B) \vee ((x,u,w,t) \in A \wedge \exists z \exists w \, (z,w,v,t) \in C) \}$$

Expanding the right hand side

$$(A(x,u,w,t) \bowtie B(y,u,v,t)) \, \unicode{x2A02} \, (A(x,u,w,t) \bowtie C(z,w,v,t))$$

we get

$$\{ (x,u,w,v,t) \mid \exists y \, ((x,u,w,t) \in A \wedge (y,u,v,t) \in B) \vee \exists z \, ((x,u,w,t) \in A \wedge (z,w,v,t) \in C) \}$$

or, equivalently

$$\{ (x,u,w,v,t) \mid ((x,u,w,t) \in A \wedge \exists y \, (y,u,v,t) \in B) \vee ((x,u,w,t) \in A \wedge \exists z \, (z,w,v,t) \in C) \}$$

The left hand and the right hand sides are identical if attributes u and w vanish.

Proof of distributivity of union over join is similar.

5. QUANTIFIERS

Quantification is an essential ingredient of predicate calculus. Informally, existential quantifier is interpreted as an infinite disjunction:

$$\exists x \, A(x,y) = A(0,y) \vee A(1,y) \vee A(2,y) \vee A(3,y) \vee \ldots$$

where we assumed that the interpretation domain for the variable x is the set of nonnegative integers. Symmetrically, universal quantifier is an infinite conjunction

$$\forall x \, A(x,y) = A(0,y) \wedge A(1,y) \wedge A(2,y) \wedge A(3,y) \wedge \ldots$$

Hence the symbols $\bigwedge_x$ and $\bigvee_x$, which are sometimes used as an alternative notation for quantifiers.

These two constructions prompt introducing the lattice *infimum*

$$\unicode{x2A02}_x A(x,y) \stackrel{\text{def}}{=} A(0,y) \, \unicode{x2A02} \, A(1,y) \, \unicode{x2A02} \, A(2,y) \, \unicode{x2A02} \, A(3,y) \, \unicode{x2A02} \, \ldots$$

where expression A(0,y) is an abbreviation

$$A(0,y) \stackrel{def}{=} [y] \bigtriangleup ( A(x,y) \bowtie [x=0])$$

for the relation A(x,y) with variable x substituted by individual constant 0. (This substitution is similar to variable renaming that we've seen in section 2). Injecting these expressions for all the constants in the domain into the lattice infimum definition and simplifying the expression by inner union associativity and idempotence we have

$$\bigtriangleup_x A(x,y) = [y] \bigtriangleup (A(x,y) \bowtie [x=0]) \bigtriangleup (A(x,y) \bowtie [x=1]) \bigtriangleup \ldots$$

In this case join distributes over union, so we get

$$\bigtriangleup_x A(x,y) = [y] \bigtriangleup (A(x,y) \bowtie ([x=0] \bigtriangleup [x=1] \bigtriangleup \ldots ))$$

Next, union of the individual constants from the domain gives the full domain relation, which could be dropped from the join

$$\bigtriangleup_x A(x,y) = [y] \bigtriangleup A(x,y)$$

Therefore, the lattice infimum is a projection.

The lattice *supremum* is defined symmetrically

$$\bowtie_x A(x,y) \stackrel{def}{=} A(0,y) \bowtie A(1,y) \bowtie A(2,y) \bowtie A(3,y) \bowtie \ldots$$

Unlike the infimum it is not domain independent. This setback indicates that perhaps the finite versions of the infimum and supremum operations might be more sucessful.

Consider a finite infimum expression

$$\bigtriangleup_{x \in B(x)} A(x,y)$$

where the variable x now ranges in an explicit set of values. Like infinite counterpart the finite infimum partitions the relation A(x,y) into a union of smaller relations with fixed attribute x each, but then selects only those partitions which have x∈B(x). This is very similar to tuple iteration semantics of SQL subquery execution

```
select distinct y from A
where x in (select x from B)
```

By repeating the steps that we performed earlier the finite infimum reduces to join and projection

$$\bigwedge_{x \in B(x)} A(x,y) = [y] \bigwedge ( A(x,y) \bowtie B(x) )$$

This is a formal derivation of subquery unnesting rule.

The finite supremum

$$\bigvee_{x \in B(x)} A(x,y)$$

turns out to be the most interesting expression of all. It is the *relational division*. Unlike infimum it can't be reduced. As it has been mentioned earlier, the relational division is exactly what is misssing in order for the lattice algebra to be relationally complete.

## 6. ACKNOWLEDGEMENTS

This subject was extensively discussed at comp.databases.theory Usenet forum. Many thanks to Jan Hidders, Jon Heggland, Val Carey, and other participants. Authors are especially grateful to the anonymous TODS reviewer whose constructive criticism of the first generation of this manuscript served as research guidance.